\documentclass[prl,twocolumn]{revtex4}

\usepackage{graphicx}

\begin{document}

\title{Pseudogap and antiferromagnetic correlations in the Hubbard model}
\author{
Alexandru \ Macridin$^{1}$,  M.\  Jarrell$^{1}$, Thomas Maier$^{2}$,  P.\ R.\ C. Kent$^{1,2}$
and Eduardo D'Azevedo$^{2}$}

\address{
$^{1}$ University of Cincinnati, Cincinnati, Ohio, 45221, USA\\
$^{2}$Oak Ridge National Laboratory, Oak Ridge, Tennessee, 37831, USA}

\date{\today}

\begin{abstract}
 
  Using the dynamical cluster approximation and quantum monte carlo we
  calculate the single-particle spectra of the Hubbard model with
  next-nearest neighbor hopping $t'$.  In the underdoped region, we find
  that the pseudogap along the zone diagonal in the electron doped
  systems is due to long range antiferromagnetic correlations.  The
  physics in the proximity of $(0,\pi)$ is dramatically influenced by
  $t'$ and determined by the short range correlations. The effect
  of $t'$ on the low energy ARPES spectra is weak except close to the
  zone edge. The short range correlations are sufficient to yield a
  pseudogap signal in the magnetic susceptibility, produce a concomitant
  gap in the single-particle spectra near $(\pi,\pi/2)$ but not necessarily
  at a location in the proximity of  Fermi surface.

\end{abstract}

\maketitle

\paragraph*{Introduction --}
The normal state phase of   high $T_c$ superconductors 
at low doping, the  pseudogap (PG) region,
is characterized by strong antiferromagnetic (AF)
correlations and a depletion of low energy states detected by both one
and two-particle measurements\cite{pgrev}.  Whereas the
d-wave superconducting phase appears to be universal in the
cuprates\cite{sc_h,sc_e}, the PG region displays different
properties in the electron and hole doped
materials\cite{h_arpes,e_arpes}.  In order to further develop a theory for
high $T_c$ superconductivity it is essential to have a better
understanding of the asymmetry and similarities between the electron and the hole doped
materials.

In the hole doped cuprates the antiferromagnetism is destroyed quickly
upon doping (persisting to $\approx 2\%$ doping)\cite{h_afm} and the
angle resolved photoemission spectra (ARPES) show well defined
quasiparticles close to ($\pi/2,\pi/2$) in the Brillouin zone (BZ) and
gap states in the proximity of ($0,\pi$)\cite{h_arpes,h_arpes1,arpes}.
In the electron doped cuprates AF is more robust (persisting to
$\approx 15\%$ doping)\cite{e_afm} and the ARPES at small doping
($\approx 5\%$) shows sharp quasiparticles at the zone edge and gap
states elsewhere in the BZ\cite{e_arpes,arpes}.  In the Hubbard model,
or the closely related t-J model, the electron-hole asymmetry can be
captured by including a finite next-nearest neighbor hopping
$t'$\cite{eh_asym,2band}.
In this Letter we employ a
reliable technique, the dynamical cluster approximation
(DCA)\cite{hettler:dca,maier:dca}, on relatively large clusters, to
investigate the PG and single-particle spectra at small doping, the
asymmetry between electron and hole-doped systems, and the role of AF
correlations on the PG physics.

We find that in the hole doped systems, the PG emerges in the proximity of
($0,\pi$), requires only short range correlations, and its magnitude
and symmetry is strongly influenced by $t'$.  In the electron doped
systems, the PG emerges along the diagonal direction, as a direct
consequence of AF scattering, and requires long range AF correlations,
but not necessarily long range order.  
The hopping $t'$ enhances the
AF correlations in the electron doped system and produces this AF
gap. With reduced temperatures, the short range AF correlations suppress 
the low-energy spin excitations in both electron and hole doped systems  
concomitant with the development of a single-particle gap in the proximity 
of $(\pi,\pi/2)$, but not necessarily with a  PG close to the nodal 
or the antinodal points.

\paragraph*{Formalism --}
The Hubbard Hamiltonian is

\begin{equation} 
\label{eq:hams}
H=-t \sum_{\langle ij\rangle, \sigma} c^\dagger_{i\sigma}c^{\phantom\dagger}_{j\sigma}
-t' \sum_{\langle\langle il\rangle \rangle, \sigma} c^\dagger_{i\sigma}
c^{\phantom\dagger}_{l\sigma} +  U\sum_i n_{i\uparrow} n_{i\downarrow}~.
\end{equation}
\noindent Here $c^{(\dagger)}_{i\sigma}$  destroys (creates) an
electron with spin $\sigma$ on site $i$ and $n_{i\sigma}$ is the
corresponding number operator. $U$ is the on-site  repulsion
taken $U=W=8t$ and  $t$ ($t'$) is the (next) nearest-neighbors hopping. 
$W$ is the electronic bandwidth.
We  keep the filling $n<1$ and take $t'=-0.3t$ ($t'=0.3t$)\cite{lda} to represent 
the  hole (electron) doped cuprates.  We present results for $n=0.95$.

In the DCA\cite{hettler:dca} we map the original lattice model onto a
periodic cluster of size $N_c=L_c\times L_c$ embedded in a
self-consistent host.  The correlations up to a range $\xi\lesssim
L_c$ are treated accurately, while the physics on longer length-scales
is described at the mean-field level.  The reduction to an effective
cluster model is achieved by coarse graining the BZ into $N_c$ cells
(see Fig.3 in Ref.\cite{maier:dca}) and approximating the self-energy as a
constant within each cell, $\Sigma({\bf k},\omega) \approx \Sigma({\bf
  K},\omega)$, where ${\bf K}$ denotes the center of the cell which
${\bf k}$ belongs to.

We solve the cluster problem using quantum Monte Carlo (QMC)
\cite{jarrell:dca3}.  We use two different 16 site cluster
geometries\cite{betts,kentbetts}, 16A and 16B (see Fig.5 in
Ref.\cite{maier:dca}), which result in different coarse graining of
the BZ. Calculations on larger clusters below the PG temperature and
at large coupling ($U = W$) are not currently possible due to the QMC
sign problem. The Maximum Entropy method\cite{jarrell:mem} is employed
to calculate the real frequency cluster Green's function from which
the self-energy is extracted. The self-energy is interpolated using a
smooth spline, and used to calculate the lattice spectrum
$A(k,\omega)$.  We find results identical to within error bars at all
the common points of the coarse-grained BZ of the
clusters. The result demonstrates that these 16 site clusters capture the
momentum dependence of the self-energy rather well.  We checked the
robustness of our results at low temperature with calculations on
smaller clusters where the sign problem is less significant.

\paragraph*{Results --}

\begin{figure}[t]
\centerline{\includegraphics*[width=3.3in,height=1.3in]{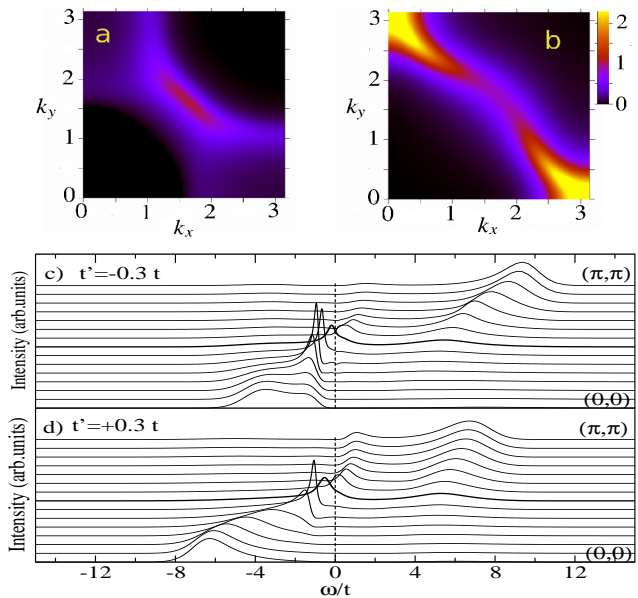}}
\centerline{\includegraphics*[width=3.3in,height=1.8in]{GMcut.eps}}
\caption{(color) $5 \%$ doping, $T=0.12t$. Zero energy surface $A(k,0)$ for a)
  $t'=-0.3 t$ and b) $t'=0.3 t$.  $A(k,\omega)$ for $k$ along the
  $(0,0)$-$(\pi,\pi)$ direction in the BZ for c) $t'=-0.3 t$ and d)
  $t'=0.3 t$.}
\label{fig:energysurf}
\end{figure}

At a temperature $T_N =0.19 t$ ($0.24 t$) for the hole (electron)
doped system the AF correlation length reaches the cluster size
yielding a divergent AF susceptibility (not shown).  Below $T_N$ one
can proceed either by imposing the full symmetry on the effective
medium, i.e. by reducing the problem to a cluster embedded in a
paramagnetic (PM) host, or by allowing the host to develop long-range
AF order.  Both the PM and the AF solutions are complementary
approximations to the exact solution: the first cuts off the AF
correlations larger than the cluster size while the second introduces
long range AF order via the mean-field character of the host.

{\emph{Paramagnetic solution -- }} In Fig.~\ref{fig:energysurf} -a and -b
we show the  spectral intensity at zero energy for the hole and
electron doped systems, respectively.  These plots are similar to the
experimental ARPES data (see Fig.8 in Ref.\cite{h_arpes} and Fig.3 in
Ref.\cite{e_arpes}).  In both experiment and in our results, a region
of large intensity can be observed close to $(\pi/2,\pi/2)$ and very
low intensity is observed at the zone edge for hole doped systems. For
the electron doped systems the intensity is maximum at $(0,\pi)$.
However, the experimental data for the electron doped materials show
gapped states along the diagonal direction \cite{e_arpes}.  In our
calculations, Fig.~\ref{fig:energysurf} -b, the intensity at
$(\pi/2,\pi/2)$ is similar to the one observed for the hole doped case
and there is no PG along the zone diagonal.  In fact $A(k,\omega)$
along the diagonal direction shows similar features for the hole
and electron doped cases (Fig.~\ref{fig:energysurf}-c and -d). Apart
from the differences at high energy close to the zone center and the
zone corner which follow the non-interacting dispersion, the
low-energy features along the zone diagonal are almost identical.

\begin{figure}[t]
\centerline{\includegraphics*[width=3.3in]{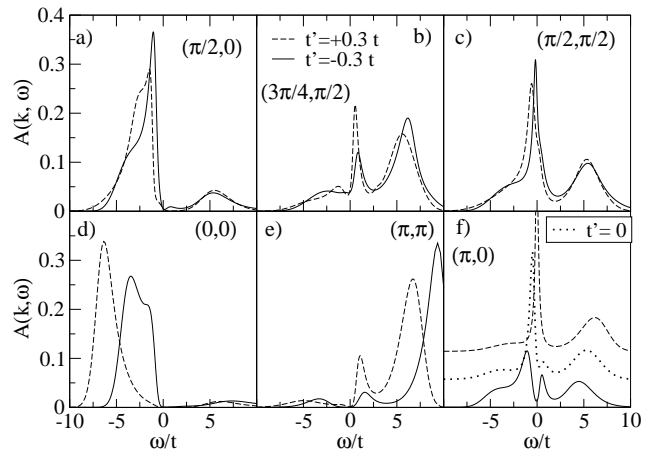}}
\caption{$A(k,\omega)$ for different $k$ points in the BZ for
hole (full line) and electron (dashed line) doped cases, at $5 \%$ doping and $T=0.12t$.  
In (f) the dotted line represents the spectrum for  $t'=0$.}
\label{fig:compeh}
\end{figure}

\begin{figure}[t]
\centerline{\includegraphics*[width=3.3in]{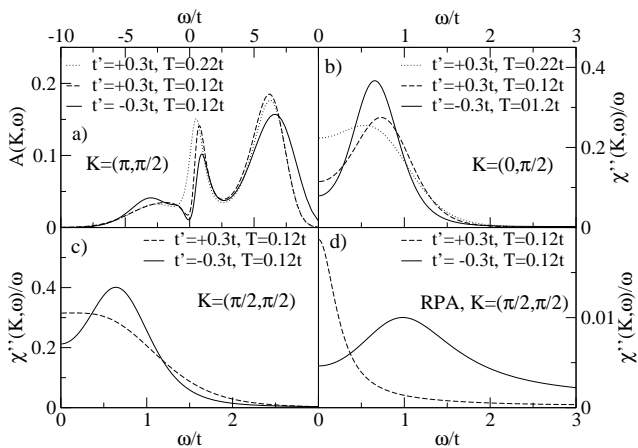}}
\caption{$5\%$ doping, hole doped (full line) and
electron doped (dashed and dotted lines) cases. a) $A(k,\omega)$ at $k=(\pi,\pi/2)$.  
b) Low-energy dynamical spin susceptibility $\chi''(q,\omega)/\omega$ 
at  $q=(0,\pi/2)$.  The suppression of the spin excitations starts
at the same temperature with the depletion of DOS at $(\pi,\pi/2)$, see dotted lines.
c)$\chi''(q,\omega)/\omega$  at  $q=(\pi/2,\pi/2)$. 
d) RPA results for $\chi''(q,\omega)/\omega$  at  $q=(\pi/2,\pi/2)$.}
\label{fig:susdos}
\end{figure}

A comparison of the hole and electron spectra is presented in
Fig.~\ref{fig:compeh} and Fig.~\ref{fig:susdos} -a, where
$A(K,\omega)$ for $K$ in the center of the cells which divide the
BZ~\footnote{The self-energy in these $K$ points results directly from
  DCA+MEM calculation. The self-energy in other  $K$ points 
  is obtained by interpolation.} are shown.  In
Figs.~\ref{fig:compeh} -a, -b, -c and Fig.~\ref{fig:susdos} -a, we
find that the single particle spectra at low energy for the hole and
the electron doped cases are surprisingly similar apart from the
features close to $(0,\pi)$. In Fig.~\ref{fig:compeh} -c we observe a
sharp peak at $(\pi/2,\pi/2)$ in both the hole and electron doped
spectra. Thus, there is no PG along the diagonal
direction~\footnote{We obtained the same conclusion from an 8 site
  cluster calculation where we could reach $T= 0.05t$ for
  $U=W$.}.  A particularly interesting feature, shown in Fig.~\ref{fig:susdos}
-a, is the depletion of the low energy states with lowering T in the
proximity of $(\pi,\pi/2)$.  Unlike the non-interacting case, where at
$(\pi,\pi/2)$, there is only spectral weight for $\omega>0$, there is
now a broad feature with substantial weight at negative energies. This is due to AF
scattering as can be deduced by comparing the main features with the
$(\pi/2,0)$ spectrum (Fig.~\ref{fig:compeh} -a) found at the mirroring
position with respect to AF zone boundary in the BZ. These shadow
states develop a gap with decreasing temperature as shown in
Fig.~\ref{fig:susdos} -a where a large temperature spectrum
($T=0.22t$, dotted line) and a low temperature one ($T=0.12t$, dashed
line) are plotted for the electron doped case.  
In literature the common description\cite{ino} 
of ARPES along the $(\pi,0)-(\pi,\pi)$ cut is that the peak at 
$(\pi,0)$ and $\omega<0$ characterizing the PG evolves into a broad feature which loses
intensity when approaching the zone corner.
Our results indicate that  the broad feature and gap at
$(\pi,\pi/2)$ are not conditioned by  the $(0,\pi)$ PG, 
being present in both electron and
hole doped systems.

Differences between the hole and electron doped spectra are
illustrated in Figs.~\ref{fig:compeh} -d through -f.  The
high energy features at  zone center (Fig~\ref{fig:compeh} -d)
and zone corner (Fig~\ref{fig:compeh} -e) are strongly influenced by 
$t'$. It is interesting that  the position of these
features seems to  follow  the  non-interacting
band structure,  the energy difference between the 
non-interacting states at these points being about $8 | t'|$. 
A fundamental difference between the electron and hole doped spectra
at $(0,\pi)$ is shown in Fig.~\ref{fig:compeh}-f. The hole doped
spectra exhibits a strong gap whereas the electron doped spectra has
an intense peak.  It is also worth looking at the $t'=0$ case, shown
in the figure with dotted line, where a PG is present but  much
less developed than the one for $t'=-0.3 t$.  Notice that even for the hole
doped case (i.e. $t'<0$) the magnitude of $t'$ has a strong influence
on symmetry with respect to zero energy of the density of states, which 
we believe to be important for interpreting tunneling experiments.

\begin{figure}[t]
\centerline{\includegraphics*[width=3.3in,height=1.7in]{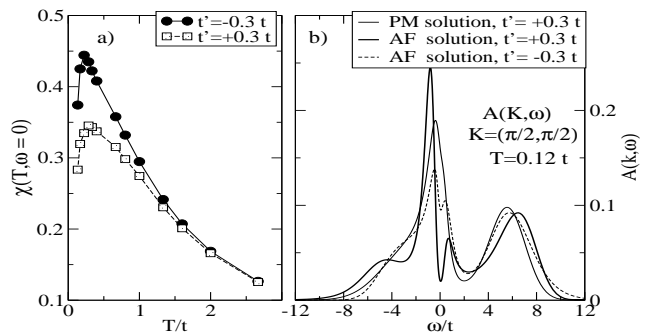}}
\caption{ a) Static spin susceptibility versus T. b) $A(k,\omega)$ averaged over the $(\pi/2,\pi/2)$ cell for the
  AF and the PM solutions.}
\label{fig:afpm}
\end{figure}

Aside from the depletion of DOS at the 
chemical potential, the PG is also associated with 
the suppression of the low-energy spin excitation\cite{spin_sus}.
For both electron and hole doped cases we find that the static
spin susceptibility  at $q=(0,0)$ is strongly  suppressed 
at temperatures below $T^{*} \approx 0.24 t$, see Fig.~\ref{fig:afpm}-a.
The momentum dependence of the dynamical spin susceptibility exhibits  
a  dispersion similar to magnon one in the undoped antiferromagnets,
with zero energy excitations at $q=(0,0)$ and $q=(\pi,\pi)$,
and gapped excitations at other $q$ points in the BZ.
For instance in Fig.~\ref{fig:susdos} -b we show the imaginary part of
the dynamical spin susceptibility, $\chi''(q,\omega)/\omega$  
at $q=(0,\pi/2)$. In the electron
doped case the position of the peak is found at  larger energy 
which can be interpreted as a larger effective exchange interaction $J$
and is consistent with  stronger AF. 
What is interesting is that the suppression of  the spin excitations  and the
formation of the remnant magnon peaks in  $\chi''(q,\omega)/\omega$ 
is concomitant with the development of the gap in the single particle 
spectrum at $(\pi,\pi/2)$ (see Fig.~\ref{fig:susdos} -a) and 
not necessarily imply a PG  at $(0,\pi)$,  observed only  in the hole doped case,
or elsewhere in the proximity of Fermi surface. However,
the  hole doped PG  at $(0,\pi)$ is also coincident with
the appearance of the remnant magnon peaks, indicating
a short range AF correlations origin.

The spin  excitation spectra are qualitatively  similar 
for the electron and hole doped cases, but with a  stronger suppression 
at low-energy  in the hole doped case even though the AF is weaker. 
The largest difference can be observed at  $q=(\pi/2,\pi/2)$ where at $T=0.12t$ the hole doped 
spectra show a well developed  gap whereas   the electron doped one  
just starts to form (see  Fig.~\ref{fig:susdos} -c). This difference
is a result of the corresponding differences in the ARPES, as can be 
concluded from Fig.~\ref{fig:susdos} -d. Here the Random Phase Approximation 
(RPA) using the calculated $A(k,\omega)$ was employed 
for the calculation of the spin susceptibility. In this approximation
the hole doped  spin susceptibility at $(\pi/2,\pi/2)$ is gapped due 
to the  gap at  $(0,\pi)$  in the DOS which suppresses the 
excitations between the antinodal and the nodal points, unlike the 
electron doped susceptibility which is peaked at low-energy.

{\em{Antiferromagnetic solution --}} In Fig.~\ref{fig:afpm} -b, we compare
$A(k,\omega)$ for AF and PM cases. Here, a gap is obtained for the AF
electron doped case close to $(\pi/2,\pi/2)$ in the BZ, in agreement
with the experimental findings\cite{e_arpes}. This gap is an
AF gap and requires long range AF correlations. The
short range AF correlations, of the order of a few lattice constants,
are not sufficient to produce it.  However, it is possible the PG
to also appear in the PM state in large enough clusters that allow
for long-range AF correlations.
This conclusion is similar to the
weak coupling PG mechanism predictions\cite{senechal}, even though 
in our case $U=W$.  The hopping $t'$ enhances the antiferromagnetism in 
the electron doped systems, producing the gap. Presumably any other parameters which favor the
antiferromagnetism will have a similar effect.  For example, the AF
solution for the hole doped case produces a gap at $(\pi/2,\pi/2)$
too, though a little smaller due to weaker antiferromagnetism.  The
spectral features away from $(\pi/2, \pi/2)$ within the AF solutions
are not qualitatively different from the ones obtained with the PM
solution (not shown). We note that the long range AF order does not
yield a gap at $(0,\pi)$ for the electron doped case even though this
point is on the AF zone boundary.
The gap in DOS at $(\pi,\pi/2)$ developed in 
the PM solution is now enhanced by AF order, as well the intensity
of the shadow states.

Our conclusions about the nature of the PG in the electron doped
systems are different from those drawn from  cluster perturbation
theory (CPT)\cite{senechal}. For  $U \approx W$, 
CPT finds that, even when only short range
AF correlations are considered, the states along the diagonal
direction develop a gap, which  persists even at large dopings
$\approx 15\%$. Whereas in the  experiment,  at this  doping value, 
the PG shows only at the intersection of the AF zone boundary 
with the non-interacting Fermi surfaces (hot spots)\cite{hot_spots}. 
For agreement with experiment
the authors of Ref.\cite{senechal} proposed two different mechanisms
for the PG in electron doped systems: a strong-coupling ($U \approx W$)
PG  at small doping  produced by  short range correlations and
a weak-coupling ($U < W$) PG valid at intermediate doping which requires 
long range AF correlations.
In contrast, we find no PG along the diagonal direction in the strong coupling regime
unless long ranged AF correlations are considered, 
implying   no need for two different
PG mechanisms in the electron doped systems.  A plausible reason for the
discrepancy between the DCA and CPT results is the overestimation of AF
correlations in the latter approach due to finite size
effects\cite{jarrell:dca3} on small
clusters\cite{tohyama} and lack of self-consistency\cite{CPT}.

We find that the inclusion in our calculation of  
a next-next-nearest-neighbor hopping $t''\approx 0.2 t$\cite{lda}
will not change the conclusions, this term having a rather small
quantitative effect (though it may provide better agreement with
experimental data).
With \emph{decreasing doping},  $T^*$ increases but the
number of available low-energy unoccupied states becomes smaller and therefore
the PG features are more difficult to be resolved.  With \emph{increasing
doping}, $T^*$ decreases and the PG weakens, its features being
hardly discernible above $15\%$ doping.  In the same time the
$\omega<0$ weight in the DOS at $(\pi,\pi/2)$ is reduced, indicating
weaker AF scattering with increasing doping.  In the AF solution the
PG at larger doping will be located at the ``hot spots'', even though
the physics at those points requires fine $k$ resolution and therefore
is obtained from interpolation.

\paragraph*{Conclusions --}
Using DCA we investigated  the Hubbard model with next-nearest neighbor
hopping $t'$. We find that: (1) The PG  along the diagonal direction of
the BZ in the electron doped systems is an AF gap which requires long
range AF correlations.
(2) The DOS in the proximity of $(0,\pi)$ is
determined by the short range AF correlations and it is strongly
influenced by $t'$. For the hole (electron) doped systems $t'$ yields a
gap (an intense peak) at the zone edge. (3) Except in the proximity of 
$(0,\pi)$ the influence of $t'$ on the low-energy ARPES  is very weak.
(4) t' has a strong influence on the high energy ARPES close to zone
center and zone corner. (5) The magnitude of $t'$  in the hole doped
systems  influences strongly the symmetry of the PG around the chemical
potential. (6) The short range AF correlations suppress the low-energy
spin susceptibility, produce remnant magnon peaks in the spin excitation
spectra in both electron and hole doped systems and produce a  gap in
ARPES around $(\pi,\pi/2)$ but not necessarily in the proximity of the
Fermi surface. When a $(0,\pi)$ PG is present in the ARPES, it emerges
at the same temperature as the magnon peaks, suggesting a common origin
of short ranged AF order. (7) Even though the antiferromagnetism is
stronger in the electron doped case, the intense peak in DOS at
$(0,\pi)$ hinders the suppression of low-energy spin excitations.

We thank George Sawatzky for useful discussions, and Kyle Shen and
Donghui Lu for sharing their ARPES data. This research was
supported  by grants NSF DMR-0312680, NSF DMR-0113574, CMSN DOE
DE-FG02-04ER46129 and  NSF  SCI-9619020 through resources provided by
the San Diego Supercomputer Center.  TM acknowledges support
from the Center for Nanophase Materials Sciences, Oak Ridge National
Laboratory, which is funded by the Division of Scientific User
Facilities, U.S. Department of Energy.

\end{document}